\documentclass[%
 twocolumn,reprint,
 amsmath,amssymb,
 aps
]{revtex4-2}

\usepackage{graphicx}
\usepackage{dcolumn}
\usepackage{bm}
\usepackage{bbm}
\usepackage{amsmath}
\usepackage{amssymb}

\begin{document}

\preprint{APS/123-QED}

\title{The Bayesian Origins of Growth Rates \\ in Stochastic Environments}
\author{Jordan T. Kemp$^1$ and Lu\'is M. A. Bettencourt$^{2,3}$}
 \affiliation{%
 $^1$Department of Physics, University of Chicago, Chicago, Illinois 60637, USA}%
\author{}%
\affiliation{%
 $^2$Department of Ecology and Evolution, University of Chicago, Chicago, Illinois 60637, USA}%
\affiliation{%
 $^3$Mansueto Institute for Urban Innovation, University of Chicago, Chicago, Illinois 60637, USA}%
\date{\today}

\begin{abstract}

Stochastic multiplicative dynamics characterize many complex natural phenomena such as selection and mutation in evolving populations, and the generation and distribution of wealth within social systems.
Population heterogeneity in stochastic growth rates has been shown to be the critical driver of diversity dynamics and of the emergence of wealth inequality over long time scales.
However, we still lack a general statistical framework that systematically explains the origins of these heterogeneities from the adaptation of agents to their environment.
In this paper, we derive population growth parameters resulting from the interaction between agents and their knowable environment, conditional on subjective signals each agent receives. 
We show that average growth rates converge, under specific conditions, to their maximal value as the mutual information between the agent's signal and the environment, and that sequential Bayesian inference is the optimal strategy for reaching this maximum. 
It follows that when all agents access the same environment using the same inference model, the learning process dynamically attenuates growth rate disparities, reversing the long-term effects of heterogeneity on inequality. 
Our approach lays the foundation for a unified general quantitative modeling of social and biological phenomena such as the dynamical effects of cooperation, and the effects of education on life history choices. 
\end{abstract}

\maketitle
Growth and inequality are fundamental properties of complex adaptive systems, and are especially important in human societies where they drive issues of prosperity and equity. 
Increased attention to these quantities and richer data enables a new set of approaches based on the statistical dynamics of populations of strategic, forward-looking agents. 
For example, we now have general answers connecting growth and redistribution schemes to specific standing levels of inequality~\cite{bouchaud,li2,stojkoski2022income}. 
However, more general questions about how societies can promote long-term growth while controlling or mitigating inequality remain rather open.  

To deal with these issues, researchers have recently sought to better understand the non-linear dynamics of wealth distributions by modeling the generation and redistribution of incomes and costs among agents within model societies ~\cite{during2008kinetic,garlaschelli,degond,chakraborti}, and by identifying resulting long-term steady-state wealth distributions  \cite{bouchaud,berman,berman2,li2,stojkoski2022income} .
In much of this work, agents representing individuals or households (often with life cycles), grow or lose wealth through a multiplicative (geometric) stochastic process. 
This modeling choice is well supported empirically and introduces a number of key parameters as an agent's resources (or wealth), $r$, evolve exponentially with mean growth rate (over time), $\gamma$, fluctuate with standard deviation (volatility), $\sigma$ \cite{bouchaud,bettencourt,patriarca} and vary across individuals of a population with standard deviation $\sigma_\gamma$ \cite{kemp2021statistical,gabaix}. 

These parameters determine the statistical dynamics of wealth in populations and the emergence of inequality across short and long timescales.  
Particularly important are the statistics of heterogeneous growth rates within the population, which generally result in growing inequality over long times. 
In such contexts, agents with higher average growth rates~\cite{kemp2021statistical} amass larger and larger amounts of relative wealth, thus reducing social mobility across the population. 
This phenomenon has been well known to economists, who have studied its emergence in models of elastic agent decision-making for goods exchanges \cite{guvenen2011macroeconomics,meghir2011earnings,blume2010heterogeneity}, and its aggregate impacts via heterogeneous growth through firm innovation  \cite{akcigit2018growth} and natural resource abundance \cite{cavalcanti2011growth}.
Generally, the sensitive trajectories of heterogeneous multiplicative growth, observed both in multiplicative growth models and in empirical data, highlight the need for theoretical developments that can explain the origins of growth rate values, volatilities, and population variances. 
More broadly, there remain analytical gaps in our understanding of how optimal agent decisions in stochastic environments contribute to disparities in growth, and what processes influence agent decisions over time and across levels of social organization. 

Recent developments in cognitive and ecological sciences provide some additional insights into the dynamics of agent behavior, suggesting that optimal decision-making and stochasticity can be understood in terms of a formal treatment of information and learning in unknown, noisy environments. 
In this vein, researchers seeking stochastic decision-making models to explore child and adolescent development \cite{ciranka2021adolescent,wu2018generalization,hertwig2004decisions} have rethought the learning process in terms of acquiring information through (active and passive) interactions with a knowable external environment.  
Similarly, ecologists have formulated natural selection, the process through which a genotype optimally leverages its environment's structure, to maximize population growth (fitness) as a (Bayesian) learning process \cite{frankV2012natural,frank2009natural,campbell2016universal,kussell2005phenotypic,bettencourt1}.  

This connection between optimal intertemporal decisions, information, and growth was originally developed as a mathematical formalism to optimize betting and portfolio investment returns~\cite{kelly,cover}. However, it has now begun to be explored as a basis for a general statistical mechanics of wealth growth and inequality~\cite{bettencourt1}.


Here we bring these insights together to develop a statistical dynamics of growth and inequality in a population of strategic agents, where the growth rates result from investing and learning in a stochastic environment.
In this approach, agents invest in sequential, stochastic environmental events based on signals they receive, and grow their wealth based on the quality of their predicted allocations.  Exploring this mechanism of information-driven growth in the context of population dynamics yields a better understanding of wealth growth and disparities. More broadly, this work adds a new dimension to the study of wealth inequality that more fundamentally links disparities between wealth, growth, and agent's subjective characteristics, such as their present knowledge and their singular experience of the environment. 


Our approach treats both resources and information as dynamically coupled quantities. 
To model information dynamics, we show that learning in the joint space of environmental states and agents' signals is developed optimally in terms of Bayesian inference, translating a maximization of predictability of environmental states into that of resource allocations and growth.  
We finish by exploring the consequences of learning a shared environment on the statistics of information and wealth, and discuss the consequences of these results for the role of general education and training on population dynamics and its potential to reverse long-term inequality \cite{kemp2021statistical}.

\section*{Theory and Modelling of Information-Based Growth}

In this section, we derive a theory of growth rates in terms of informational quantities.  Here, information means an agent's predictive knowledge of event probabilities in a noisy environment.
Agents seek to maximize the growth of their resources over time by investing in a set of possible events in their environment using their individual knowledge.
This agent's knowledge is subjective in the sense that it is formed by the agent's own observations and is not shared or compared with other agents. This knowledge is improved over time through an iterative process of (Bayesian) learning based on the agent's expectations (``beliefs") and observed environmental outcomes.
We illustrate these dynamics using a multinomial model, for which we derive closed-form expressions for the average resource growth rate and volatility in terms of information-theoretic quantities.
We will then show when this learning process dynamically attenuates inequality in resource growth rates across populations.


\begin{figure*}
\quad
\includegraphics[width=.95\textwidth]{ 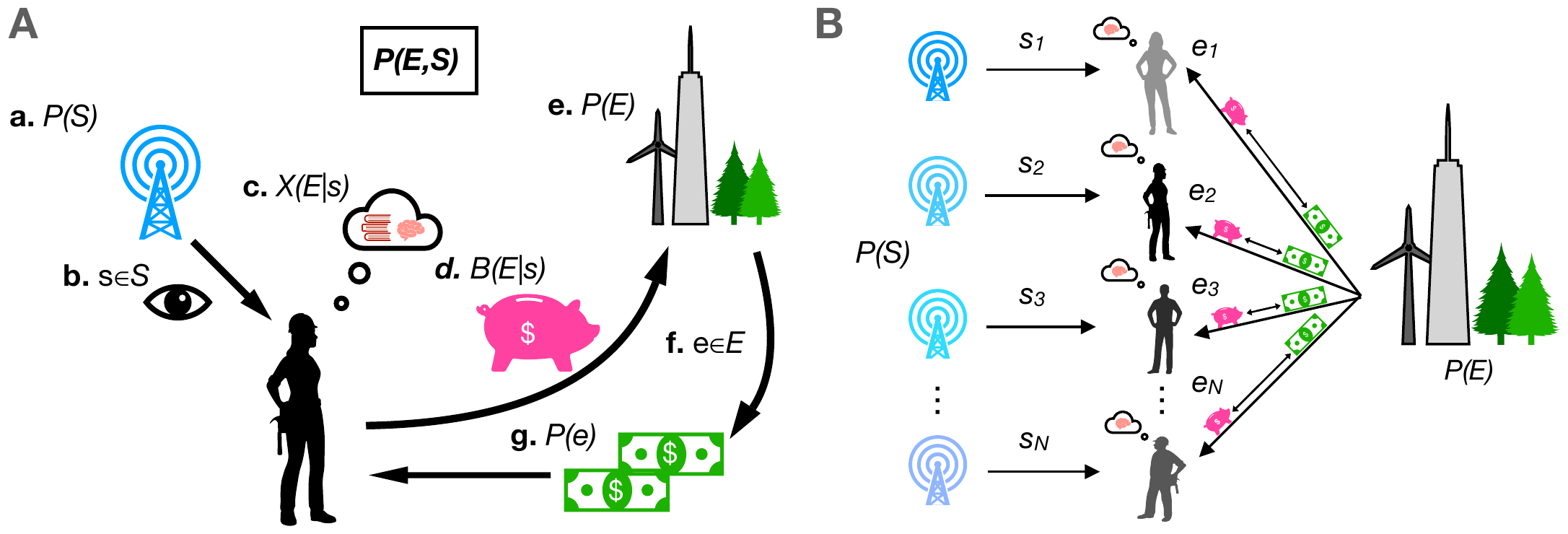} \\
\caption{\label{fig:Fig1} Agents earn resources from their environment based on the quality of their information. \textbf{A.} During each time step, \textbf{a}. the agent's private channel outputs a signal $s\in S$ with probability $P(s)$. \textbf{b}. The agent observes $s$, \textbf{c}. the agent consults their guess for the  conditional outcome probability of the environment, $X(E|s)$, and \textbf{d.} the agent makes proportional bets on all outcomes $B(E|s)$. \textbf{f,e.} The true event $e\in E$ is revealed from the environment with probability $P(e)$, and \textbf{g.} the agent receives a payout proportional to the marginal probability of $e$. \textbf{B.} In a population simulation, $N$ agents independently sample private signals and invest in events sampled from the same environment.}
\end{figure*}

\subsection*{Growth from Information}

We consider a population of $N$ agents, each initially assigned resources $r_i$ that can be (re)invested into the set of outcomes of their environment to generate returns.  The agents have access to a private predictive signal $s\in S$, which they use as a predictor to invest resources in events $e\in E$ generated by their environment.
The set of signals and events are described by the joint probability distribution, $P(E,S)$ with marginals $P(E)$ and $P(S)$.  

At each time step, each agent observes its own  signal $s$, and allocates its resources on events, using a vector $B(E|s)$.
As the event $e$ is revealed, the agent is awarded returns, $w_e$ for the fraction of resources invested in the correct outcome, $B(e|s)r_i$. 
After $n$ steps, the agent's total resources (wealth) is

\begin{equation}\label{jordanbets}
r_{n}=r_i\prod_{j=1}^n B(e_j|s_j)w_{e_j}=r_i\prod_{s,e}\big[B(e|s)w_e\big]^{W_{s,e}},
\end{equation}

\noindent where $W_{s,e}$ is the number of occurrences (wins) of $s,e$. 
Note that $\frac{W_{s,e}}{n}\rightarrow P(s,e)$ as $n\rightarrow\infty$ by the law of large numbers. It follows that the average growth rate of resources over $n$ steps is 

\begin{equation}\label{eq:growthrate}
    \gamma_i\equiv\frac{1}{n}\log \frac{r_n}{r_i}\approx\sum_{e,s}P(s,e)\log[B(e|s)w_e].
\end{equation}

Kelly showed that the maximal growth rate as $n \rightarrow \infty$, obtained by maximizing the previous expression with relation to $B(E|S)$, results in an allocation mirroring the conditional probability, $B(E|S)=P(E|S)$.  This maximum growth rate is the mutual information, $\gamma_{max}=I(E,S)$ when the odds are "fair", $w_e=1/P(e)$ \cite{kelly}. 

In reality, agents do not start out with perfect knowledge. In this case, agents must invest using their present best estimate for the conditional probability, $X(E|S)\neq P(E|S)$. Then, their resource growth rate will be smaller than the maximum. This can still be written in terms of informational quantities as the Kelly growth rate \ref{kellyinfo},

\begin{equation}\label{KGR}
    \gamma=I(E;S)-\mathrm{E}_{s}\big(D_{KL}\big[P(E|s)||X(E|s)\big]\big).
\end{equation}
\noindent where $\textrm{E}_s$ is an expectation value over the states of the signal, and $D_{KL}\big[P(E|s)||X(E|s)\big]=\sum_e P(e|s)\log\frac{P(e|s)}{X(e|s)}$ is the Kullback-Leibler divergence, expressing how similar the two distributions in its inputs are.
This result shows that agents with better information will experience greater resource growth rates, as long as they invest optimally \cite{algoet1988asymptotic}.  These compounding dynamics are illustrated in Fig. \ref{fig:Fig1}.

We will now illustrate these general results using a specific multinomial model, which will allow us to further explore population dynamics. 

\subsubsection*{Multinomial Choice Model of Growth Dynamics}


Consider the space of signals, $S$ and events $E$, of equal size $l$ with  outcomes $s,e\in\mathbb{Z}$ and degenerate, multinomial conditional probability

\begin{equation}\label{app:ppos}
    P(e|s)= f(p,l)=   
    \begin{cases}
        p & \text{if } s=e\\
        \frac{1-p}{l-1}& \text{if } s\neq e,
    \end{cases}
\end{equation}
\noindent  where $0< p< 1$ is the binomial probability of guessing the correct outcome. For simplicity, we assumed that the probability of a correct guess independent of $l$. The distribution has uniform marginals, $P(e)=1/l$ and $P(s)=1/l$, for all signals and events, such that $P(s|e)=P(e|s)$ via Bayes' rule. 

The mutual information is then $I(E;S)=\log l+p\log p+(1-p)\log\frac{1-p}{l-1}$ (APP \ref{appinf}).
For a binary choice, $l=2$, the first term gives 1 bit as the entropy of the environment and the remaining is the conditional entropy, expressing how well an agent could know the environment given the signal.
In the limit $p\rightarrow 1$, agents have perfect knowledge of the marginal of $E$.

So far we considered that the agent has perfect knowledge of the joint distribution of the signals and the environment. 
When this is not the case, we can write a parametric expression of the agent's ignorance in terms of an estimated binomial probability $x \neq p$.
The agent's likelihood model of the conditional probability is then $X(e|s)= f(x,l)$. The divergence term of Eqn. \ref{KGR} becomes the divergence between $f(p,l)$ and $f(x,l)$ averaged over all signals, $\textrm{E}_s\big[D_{KL}\big]=p\log \frac p {x}+(1-p)\log\frac{1-p}{1-x}$.
Subtracting the mutual information by this term yields the agent's growth rate in parametric form under imperfect information as (APP \ref{simplegrowth})

\begin{equation}
    \gamma=\log l+p\log x+(1-p)\log\frac{1-x}{l-1}.
\end{equation}
\begin{figure*}
\vspace{-.4cm}
\quad
\centering
    \hspace{-.3cm}\includegraphics[width=\textwidth]{ 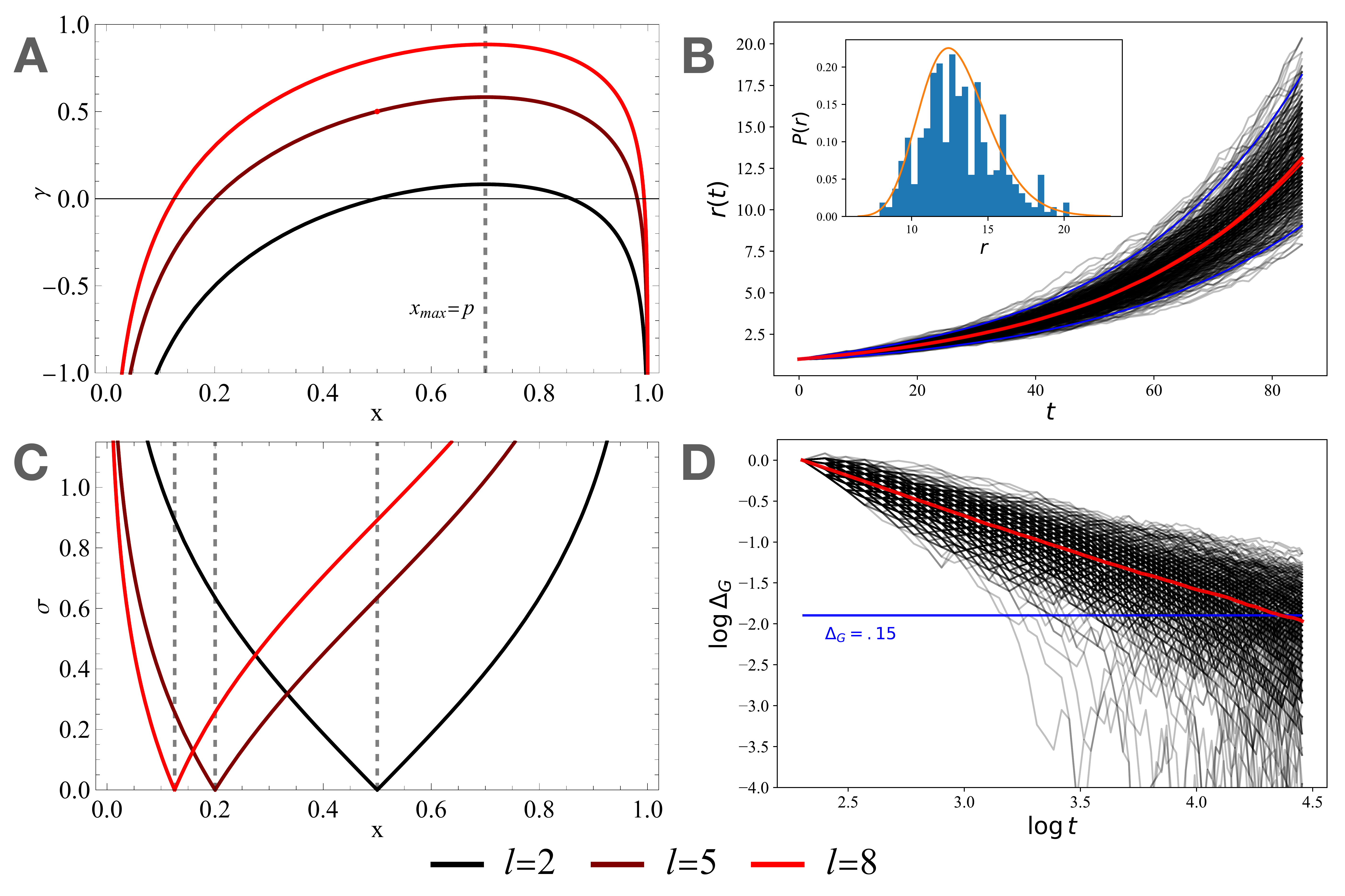}
\caption{\label{fig:Fig2} The degenerate multinomial environmental structure permits the parameterization of growth rates and volatilities, and can reproduce the behavior of GBM wealth generation models. \textbf{A}. For $p=.7$, the computed growth rate maximizes at $x=.7$, decreases as $x$ diverges from $p$, and scales with the value of $l$. The parameter $l=2$ provides a realistic range of growth rates. 
\textbf{B}. Monte Carlo simulations with $N=388$ homogeneous agents, all with $\gamma=.03$ and $r_0=1$. The population resource statistics produce lognormal exponential growth. The expected mean (red) and actual mean (orange) overlap in value. The blue lines represent the 95\% confidence interval. \textit{Inset}: The resource histogram is fit to a log-normal distribution of the same growth and volatility parameters.
\textbf{C}. Volatility minimizes at $x=l/2$ and increases monotonically in either direction. Volatility increases more rapidly at higher values of $l$. \textbf{D}.  Over time, $\Delta_\gamma\rightarrow 0$ as agents' growth rates approach the Kelly growth rate. The average agent converges to within 15\% the expected mean at $t\approx 80$. }
\end{figure*}

\noindent This expression is plotted in Fig. \ref{fig:Fig2}A as a function of $x$ for various $l$ values and fixed $p$.  We see that increasing the size of the event space, $l$, reduces the probability of any individual outcome, making it harder to guess, increasing the payouts and Kelly growth rate. The maximal growth rate is obtained when $E_e [ D_{KL}] \rightarrow 0$, when $x\rightarrow p$. Conversely, $\gamma \rightarrow 0$ when $p\rightarrow 1/l$, indicating the signal and the environment have become statistically independent. 

Treating $\gamma$ as the expected growth rate, the volatility is calculated as the second moment of the growth process. The volatility squared (variance) is given as (APP \ref{simplevariance})

\begin{equation}
\sigma^2={p(1-p)}\log^2\frac{x(l-1)}{1-x}.
\end{equation}

\noindent This expression is shown in Fig. \ref{fig:Fig2}C. The  volatility vanishes in the limit $x\rightarrow 1/l$, corresponding to when agents invest with equal probability in all possible event types. A larger $l$ increases the magnitude of the growth rate, but also the volatility.  The volatility is highest when $p\rightarrow 1/2$ and the environment is most uncertain, but the agents feel sure of the outcomes when $x\rightarrow 0$  or $x\rightarrow 1$.

Kelly's formulation describes the average growth rate of resources over a large number of discrete  investments~\cite{kelly}. To relate this situation to a growth process in time, the agent makes $\omega$ bets per unit time such that $\Delta t=1/\omega$ is the average interval of time between investments. Resources at time $t$ are then the compound of all investments made in the time interval $[t,t+\Delta t]$.
In the continuous limit, $\omega\rightarrow\infty$,  $r_n\rightarrow r(t)$ and $\gamma$ describes the average growth rate over long times. We consider $t\approx 10^{-2} yr$  so that our results are comparable to previous work based on yearly growth rates of the order of a few percent. Volatility is reduced $\sigma_t=\sigma_n/\sqrt\omega$ as fluctuations are averaged out in each time step (APP A3). 


Fig.  \ref{fig:Fig2}C demonstrates the two investment regimes for each value of $\gamma$; where the growth rate maps to either high or low volatility depending on the value of $x$.
Using $x>p$ results in over-investment, denoted as aggressive betting, as agents overestimate the dependence between their signal and the environment.  In this situation, the agent invests relatively more on diagonal outcomes and experiences large gains or losses  resulting in higher volatility.
With $x<p$, or in the conservative regime, agents underestimate $p$  and distribute their wealth more equally across all outcomes, resulting in less volatility.    Agents can also experience $\gamma=0$ at two values of $x$. In the trivial limit, as $x\rightarrow 1/l$, signals and agent investments become statistically independent. The other trivial case can be solved for numerically when $\gamma=0$. 

The dynamics of this model closely resemble the well-known behavior of geometric Brownian motion (GBM) with drift. 
Fig \ref{fig:Fig2}B shows the dynamics of a population of agents with homogeneous parameters evolved using a Monte-Carlo simulation. 
Over time, mean population resources grows with $\langle r(t)\rangle=\frac{1}{N}\sum_i r_i(t)=\exp[\gamma t]$, in agreement with \cite{bettencourt}. 
We also demonstrate that the time-averaged growth rate of resources converges to the Kelly growth rate over long times.
Fig \ref{fig:Fig2}D shows the asymptotic convergence of the normalized difference of averaged growth rate for individual agents $\Delta_G=(\gamma-G)/\gamma\rightarrow0$, where  $G=\frac{1}{t}\ln\frac{r(t)}{r(0)}$ (black) and population-averaged growth rate, $\langle G\rangle=\frac{1}{N}\sum_i G_i$ (red).

So far we have considered $x$ as a static variable and explored the dynamics of resources when $x\neq p$. We are now ready to consider $x$ as a dynamical variable that converges to $p$ as a result of sequential (Bayesian) learning. 

\subsection*{Bayesian Dynamical Growth}

More realistic agent trajectories are dynamical in the sense that investment allocations become history-dependent and reflect the cumulative knowledge of each agent's past experience  \cite{bettencourt,bayer2005midbrain}.  
In such a setting, agents can improve their information by updating their model of the conditional relationship of $S|E$ with each observation. 
In the absence of other processes, this learning task is optimally achieved in terms of sequential Bayesian inference \cite{behrens2007learning,cox1946probability}:
\begin{equation}
X_n(e|s)=A P(s_n|e_n)X(e_{n})=\left[ \Pi_{i=1}^n \frac{P(s_i|e_i)}{P(s_i)}\right] X(e),
\end{equation}
where the normalization $A=\big(\int d e_n P(s_n|e_n)X(e_n)\big)^{-1}$, and where we take the prior $X(e_1)=X(e)$, as we are assuming that the environment is stationary. 








Bayesian inference converges $X(E|S)\rightarrow P(E|S)$, maximizing the growth rate in the long run.  
The agent's interactions with the environment are thus not only a way to gather resources in the short term, but also information~\cite{bettencourt1}, as demonstrated in Fig. \ref{fig:learndiag}A. In minimizing information divergence, the  learning process  maximizes resource growth over the long term. 
In the following section, we describe a parametric Bayesian inference scheme applied to the multinomial model of the previous section, via a Dirichlet prescription of conjugate priors \cite{blei2003latent}.

\subsubsection*{Bayesian Inference in the Multinomial Model}

\begin{figure}[ht]
\centering
\quad
 \includegraphics[width=.48\textwidth]{ 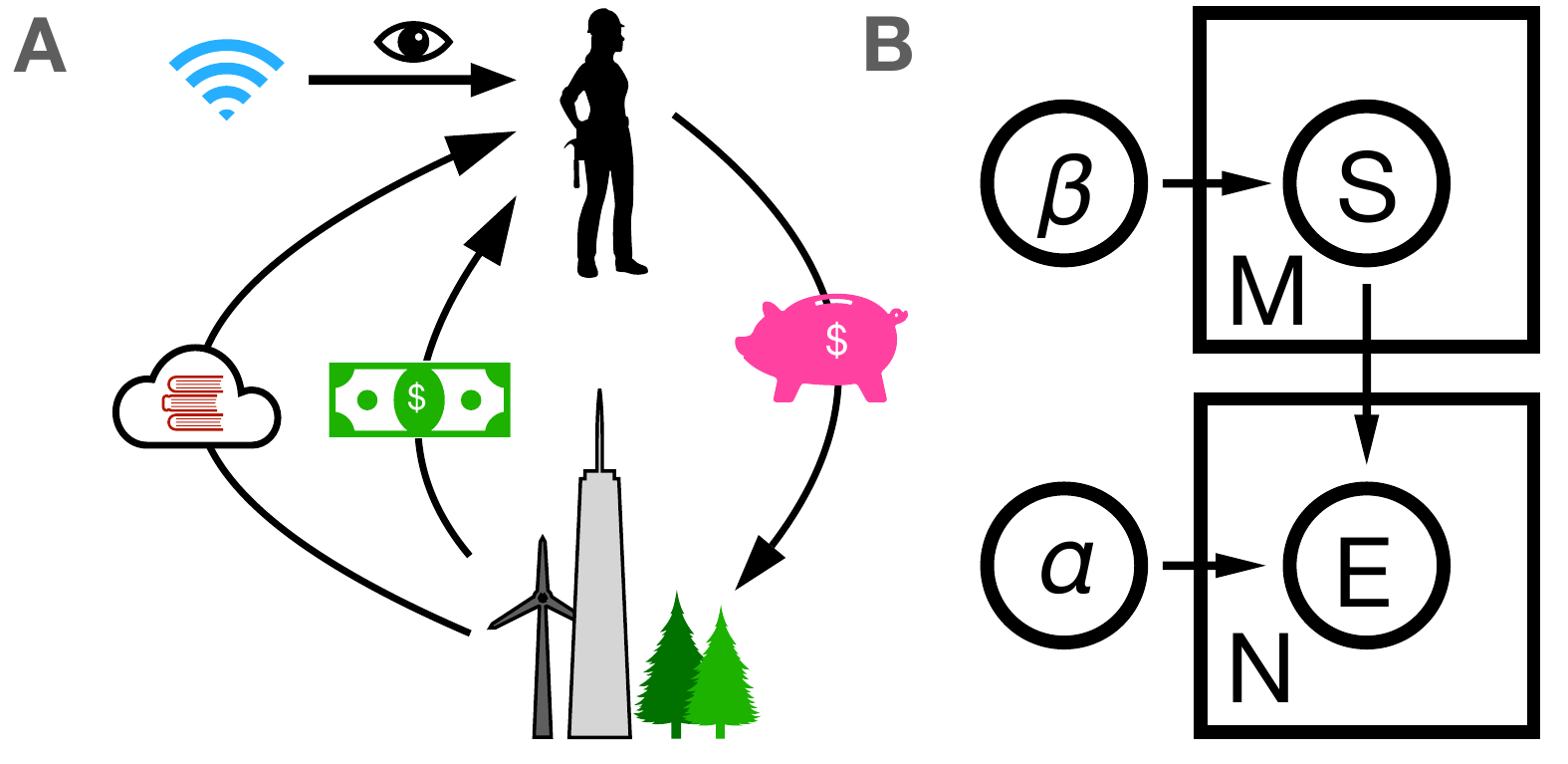} \\

\caption{\label{fig:learndiag} Visualization of the learning process. 
\textbf{A}. In addition to earning resources from the environment, the agent is now awarded information with each investment. 
\textbf{B}. Plate notation for the latent Dirichlet inference process.
The agent is assigned prior parameters $\tilde{\mathbf\alpha},\tilde\beta$, corresponding to their guess for the distributions of $E$ and $S$, which are updated based on counts $M,n$ respectively.}
\end{figure}

In general, we define the agent's likelihood of a sample of the signal, $s|e$, as a categorical distribution with parameter vector $\beta=\{\boldsymbol{\beta}^1,...,\boldsymbol{\beta}^{l}\}\in  \mathbb{R}^{l}$, with each vector corresponding to an event and each component, $\beta^e_s$ corresponding to a signal, event pair.
The probability mass function is given by  $P(s|e)=\prod_s(\beta_s^e)^s$, with normalization $\sum_s\beta_s^e=1$.
The conjugate prior distribution of $E$ is given by a Dirichlet with hyperprior vector $\boldsymbol\alpha\in\mathbb{R}^{l}$, and distribution $P(e)=\alpha_e/A$, where magnitude $A=\sum_e^l\alpha_e/l$.
This scheme is illustrated in Fig. \ref{fig:learndiag}B.

We set $\alpha_e=1$ for all $e$ so that our prior is uniform.
We ensure the off-diagonal degenerate condition by setting $\beta^e_s=p_e$ for $s=e$, and for off-diagonal events, $e\neq s$,  $\beta^e_s=\frac{1-p_e}{l-1}$, satisfying Eqn \ref{ppos}.
Under this setup, the binomial parameter describing the environment is given by the average along the diagonal,

\begin{equation}\label{pdyn}
\begin{split}
p&=\frac{1}{l}\sum_{s}^{l}\beta_s^e, \hspace{.5cm}s=e. \\
\end{split}
\end{equation}

We have thus far described the parameter configuration for an agent with perfect information.  An agent with imperfect information will have estimates for the parameters, $\tilde{\boldsymbol{\alpha}}\neq\boldsymbol{\alpha}$ and $\tilde \beta\neq \beta$, and posterior, $X(E|S,\tilde \beta,\tilde{\boldsymbol{\alpha}})\neq P(E|S,\beta,\boldsymbol{\alpha})$.
With each observation, the agent  updates $X(E|S)$ via (APP \ref{LDA}) 

\begin{equation}\label{eq:update}
    X(e|s)\propto \frac{m_{(-s)}^{(-e)}/\omega k+\tilde\beta_s^e}{M^{(-s)}/\omega k+1}(n_{(-e)}+\tilde\alpha_e),
\end{equation}

\noindent where $m^{(-e)}_{(-s)}$ and $n_{(-e)}$  are the cumulative number of observations of $e,s$ pairs and $e$ excluding the current observation, and $M^{(-s)}=\sum_em^{(-e)}_{(-s)}$ is the total occurrences of $s$ excluding the current.
We also introduce an inference time, $k$, as a free parameter that weighs the evidence versus the prior, with units $time/update$ such that $t/k$ is unit-less. 
In the limit $k\rightarrow \infty$, the agent does not update their prior with new evidence. In the opposite limit, $k\rightarrow 0$, the agent ignores the prior and considers only the most recent evidence, and this becomes a maximum likelihood model.

During the inference process, the agent will break the degeneracy of their posterior as they infer each $\beta^e_s$ individually.
This is inconsequential though, as $x(t)$ can still be computed similarly to equation \ref{pdyn} at any time. 
The degeneracy of $P(E|S)$ permits us to reduce the dynamics of $X(E|S)$ to that of the diagonal probability $x(t)$, such that (APP \ref{asympt}).

\begin{equation}\label{simplemeanlearn}
    x(t)= \frac{pt/kl+x_0}{1+t/kl},
\end{equation}

\noindent  where $x_0$ is the agent's initial binomial parameter.
This equation is a core result of this work, as the dynamics of the information stored in the agent's posterior determine the average dynamics of the growth rate, via the functional $\gamma[x(t)]$.
Over many observations, the agent refines their guess, driving $X\rightarrow P$, minimizing their information divergence as $D_{KL}(P||X)\rightarrow 0$.
The agent thus maximizes the growth rate for their signal over time with a power law $-1$ in terms of the unitless inference parameter $\lambda\equiv t/kl$.
For the remainder of this paper, we will study the effects of this learning process on the population dynamics of growth rates and wealth.

\subsection*{Population effects of Bayesian Dynamics}

Having defined the dynamics of single agents, we can now explore the dynamics of growth rate statistics in a heterogeneous population.  Growth rates can vary because agents have different initial conditions, experience varying stochastic histories, or have different inference models (likelihoods).

To better understand the effects of these sources of heterogeneity, we write the population variance of growth rates in terms of information-theoretic quantities.
Where for convenience, $I_i\equiv I(E;S_i)$ and $D_i\equiv \textrm{E}_{s_i}\big(D_{KL}\big[P(E|s_i)||X(E|s_i)\big]\big)$, the population variance is given as (APP \ref{GRV})

\begin{equation}    \textrm{Var}_N[\gamma_i]=\textrm{Var}_N[I_i]+\textrm{Var}_N[D_i]-2\textrm{Covar}_N\big[I_iD_i\big].
\end{equation}
\noindent The first two terms in this equation arise from different sources and have different implications for inequality.
The first term is independent of the learning process and trajectory and depends only on the model of the environment given the agent's signal.
The second term expresses variance in the prior and learning trajectory, and eventually vanishes as agents learn their environment fully.
Thus, the population growth rate variance  only vanishes if every agent has a signal with the same statistics, and after every agent has had time to learn their environment.
The third term arises in populations where the quality of signals correlates with agents' information on the signal, a notion particularly relevant in modeling environments with high variability across signal types. 
For example, across the United States towns of different sizes, growth rates and wealth positively covary with inequality as cities are often both wealthier and more unequal wages than smaller municipalities \cite{heinrich2021scaling,
bettencourt2007growth}.
In this work, we focus on the inference process for agents with identically distributed signals.
That is we (implicitly) take $\textrm{Var}_N[I_i]=0$ and ignore any covariances.

For a population of agents in the multinomial environment with heterogeneous information independently sampling the same signal, the initial variance in growth rates is given by the variance in the initial binomial parameter, $\sigma^2_x$. The dynamics of the variance in binomial parameter for a population of size $N$ is (APP 5.1)
 
\begin{equation}\label{xvar}
\textrm{Var}_N\big[x_i(t)\big]\equiv \big\langle\big[x_i(t)-\langle x(t)\rangle\big]^2\big\rangle=\frac{\sigma_x^2}{(1+t/kl)^2},
\end{equation}

\begin{figure*}
\vspace{-.4cm}
\quad
\centering
    \hspace{-.3cm}\includegraphics[width=\textwidth]{ 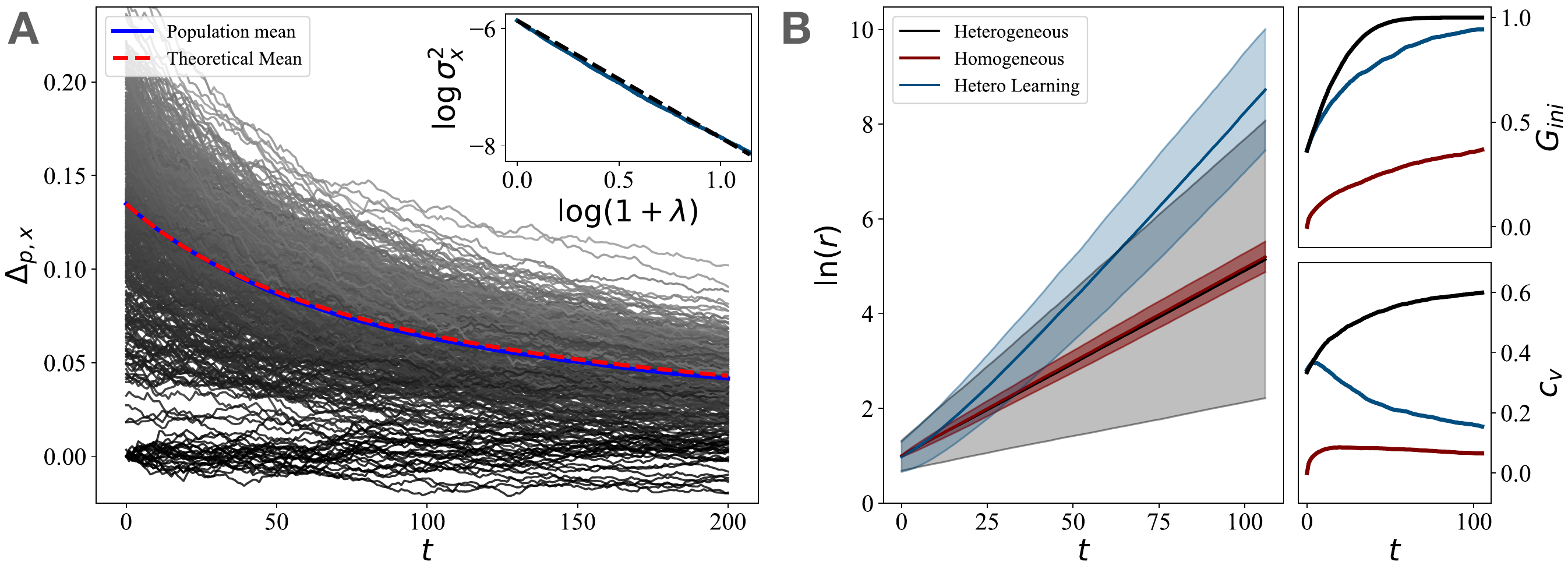}
\caption{\label{fig:Fig4}  Monte Carlo simulations of a population undergoing the inference process with parameters of mean growth rate, $\bar \gamma=.04$, and standard deviation, $\sigma_\gamma=.641\bar\gamma$. \textbf{A}. 
The population and theoretical means of $x$ converge to $p$, thus maximizing growth rate.
The parametric variance, $\sigma^2_x$, (blue) follows the theoretical prediction (red). 
The linear behavior log-log plot demonstrates the power law behavior of $\sigma^2_x$.
\textbf{B.} The mean resources of three population types are plotted with a shaded region providing $95\%$ CI bounds for single agent trajectories. Heterogeneity broadens the range of possible wealth values, while inference both increases mean growth while narrowing the shaded region relative to no inference. Agent inference slows the increase in the Gini coefficient introduced by heterogeneity, and reduces the coefficient of variation.}
\end{figure*}
 
\noindent where $\langle x(t)\rangle=\frac 1N\sum_ix_i(t)$, and $\sigma_x$. 
Assuming a population of entirely conservative (or aggressive) agents, such that all growth rates map to a unique binomial parameter, we can approximate the variance in growth rates,  
$\sigma^2_\gamma(t)=\big\langle\big(\gamma[x_i(t)]-\gamma[\langle x(t)\rangle]\big)^2\big\rangle$, by Taylor expanding the second moment of the resource distribution.
APP \ref{growthVar} shows that variance decreases asymptotically in polynomial $t^{-2}$ time. 
Figure \ref{fig:Fig4}A demonstrates that in a population of agents sampled from a Gaussian distribution of growth rates and resources learning their environment, $\Delta_{p,x}=p-x(t)\rightarrow0$ as $t\rightarrow\infty$, and their binomial parameters converge to the optimal. 
On the population level, there is agreement between the empirical population mean and theoretical mean trajectory, calculated by evolving $\langle x(t)\rangle$ using Eqn \ref{simplemeanlearn}.
Similarly, the empirical population variance in $x$ matches the theoretical power law prediction given by Eqn \ref{xvar}.

This result shows that learning a shared environment reduces growth rate variance on the same time scale as the dynamical effects introduced by growth rate variance \cite{kemp2021statistical}. 
It suggests that fast learning (by a sufficiently low $k$) equalizes information access, and is a suitable mechanism for reversing the long-term effects of heterogeneous growth on inequality. 
We demonstrate this by comparing the statistics of resources across Monte Carlo-simulated populations; first with homogeneous initial conditions, then with heterogeneous initial conditions with and without inference. 
To measure the increase in inequality, we track the Gini coefficient, denoted $G_{ini}$, a value that scales from 0, for uniformly distributed resources, to 1, for maximally unequal resources.
In a distribution that is lognormal in the continuous limit such as in these simulations, $G_{ini}(t)\approx \textrm{Erf}[\sigma_r^2(t)]$.
We furthermore measure the relative increase in variance to resources using the coefficient of variation $c_v=\sigma_r/\langle r\rangle$, to assess whether the increase in inequality outpaces the overall resource growth. 
More on this analysis is given in \cite{kemp2021statistical}.

The resource time series in Figure \ref{fig:Fig4}B demonstrates that growth rate heterogeneity dramatically broadens the wealth distribution, in agreement with \cite{kemp2021statistical}.
Accordingly, heterogeneity increases $G_{ini}$ and $c_v$ as compared to a homogeneous population. 
The introduction of learning increases the average growth rate in a heterogeneous population, as demonstrated by the higher mean wealth, while reducing the variance in resources. 
The former slows the rapid increase $G_{ini}$, while the combination of both reduces $c_v$ to levels comparable to the homogeneous trajectory, confirming that learning reverses the effects of heterogeneity on inequality.

While this simplified model does not capture the nuanced effects of educational systems in real societies, the connection between convergent learning in a population and growth is general and provides a sound theoretical basis for the observed benefit of education on national growth, human capital, and inequality reduction \cite{morris1996asia,krueger2001education,hanushek2010education}.

\section*{Discussion} 
We developed a statistical theory for the origin of resource growth rates in populations of learning agents experiencing a correlated stochastic environment. We showed that an agent's growth rate is, in the long time limit, the quantity of mutual information they possess about their environment, and that learning through Bayesian inference provides a natural (optimal) mechanism for increasing agent's growth rates, managing volatility, and reducing disparities across populations over time.
We demonstrated that with relatively simple modeling assumptions, this theory produces similar behaviors as GBM models widely used in studies of wealth dynamics and inequality.
The present treatment answers an important open question on how to mechanistically control variances in growth rates across a society while maximizing learning and growth, and generally enriches the typical modeling schema of wealth dynamics by incorporating agents' subjective choices in a structured statistical environment. Beyond these results, 
this work provides a modular foundation for incorporating information and strategic subjective agent behavior in statistical mechanics, bridging a gap between physics and computer science, and biological and social science. 

There are a number of important developments that this type of theory suggests and that will be necessary to model realistic social systems.  First, learning is not uniform across populations or time, varying across the life course, with some agents being able to dedicate more time and effort to it. This issue can be modeled by making inference rates dynamic and heterogeneous, for example, through coupling to agents' socioeconomic status (SES) or age. Importantly, SES has been shown to correlate negatively with the presence of stressors that inhibit the cognitive ability of people to learn \cite{weissman2021,evans2004environment,hackman2010socioeconomic}, and positively with educational outcomes \cite{braga2017wealth,lovenheim2011effect,belley2007changing}. 
Coupling learning rates to SES would alter the population learning trajectory and potentially attenuate its effectiveness in reducing information and wealth inequality.  Moreover, our analysis has assumed that each agent samples identically distributed signals.  In reality, people across different societies, cities, or even neighborhoods have access to different signals, with implications for what they can learn and for resulting social equity. Future studies of the origins of inequality and social equity should consider these structural complexities from the general point of view of access to information and learning.

Second, from the point of view of maximizing resources, there are familiar trade-offs between learning and investing. These can be modeled in terms of the inference process, which can be divided into passive experiential learning, resembling the ``learning by doing" featured above, and, additionally, emulating formal, institutional education wherein agents sacrifice short-term gains in income to more rapidly increase their knowledge and learning rates.  These considerations define agent trade-offs between actively exploring and passively exploiting the environment, an important topic in experimental neuroscience and machine learning \cite{kidd2015psychology,thrun1995exploration}. 
Furthermore, while information is a non-rival good that can be made available to a society with minimal cost or degradation, the generation and dissemination of information through teaching is a costly process that can produce additional non-trivial dynamics.  Incorporating the social costs of education through mechanisms of finite learning resources could help explore trade-offs in investing in human capital over various timescales \cite{schultz1971investment,paulsen2001economics} and determining when they are worth it in inter-temporal settings. 

Thirdly, tracking individual agent dynamics under constraints of finite (varying) lifespans can determine the effects of generational wealth transfers on inequality, and provide insight into life-course strategies~\cite{elder2003emergence} and issues of valuing the future. Thus, an extended framework can help us explore the scope of education under the discounting of delayed resources by longevity and lived volatility \cite{hannagan2015income}; including the implications of costs and expected earnings with or without an education over time.  
Lastly, agents in this model experience the same environment and learn the same information, whereas actual communities specialize in different, complementary skills that minimize knowledge redundancy. These complementarities and exchanges are known commonly in the social and ecological sciences in terms of the division of labor and knowledge in societies. How agents decide which information to learn and what profession to choose based on their environments begets different growth rates across a population, altering emerging inequality and influencing how social groups cooperate or compete across community or institutional social levels \cite{frank2012naturalIII}. Cooperation among agents with synergistic information in a stochastic environment has been shown to produce non-linear additive effects on aggregate information, suggesting that cooperative agents would experience larger growth rates when coordinated, compared to the sum of agents acting independently~\cite{bettencourt2009rules,queller1985kinship}. Studying this connection between social behavior and growth from the point of view of information and learning will provide insights into the circumstances when cooperative and altruistic behavior becomes favored from the point of view of both shared resources and information. 
 
\addcontentsline{toc}{sectio n}{Acknowledgement}
We thank Arvind Murugan, Marc Berman, and Adam Kline for their discussions and comments on the manuscript. This work is supported by the Mansueto Institute for Urban Innovation and the Department of Physics at the University of Chicago and by a National Science Foundation Graduate Research Fellowship (Grant No. DGE 1746045 to JTK).

\bibliographystyle{ieeetr}
\bibliography{refs} 

\appendix
\section{Information quantities}

\subsection{Kelly growth rate}\label{kellyinfo}

Applying $P(e|s)/P(e|s)$ to the log of Eqn. \ref{eq:growthrate} yields 

\begin{equation}
\begin{split}
    \gamma&=\sum_{e,s}P(e,s)\log\bigg[w_eP(e|s)\frac{ X(e|s)}{P(e|s)}\bigg]\\
    &=\sum_{e,s}P(e,s)\log\frac{P(e|s)}{P(e)}- P(s)P(e|s) \log \frac{P(e|s)}{X(e|s)}\\
    &=I(E;S)-\mathrm{E}_{s}\big(D_{KL}\big[P(E|s)||X(E|s)\big]\big),
\end{split}
\end{equation}

where $\textrm{E}$ denotes an expectation value over all sample outcome states.

\subsection{Simplified growth model}\label{simplegrowth}
Consider a conditional probability that is degenerate off-diagonal,

\begin{equation}\label{ppos}
    P(e|s)= f(p,l)=   
    \begin{cases}
        p & \text{if } s=e\\
        \frac{1-p}{l-1}& \text{if } s\neq e,
    \end{cases}
\end{equation}

The ``correct" outcome corresponding to the sampled event occurs with conditional probability $0<p\leq1$, and all other "incorrect" guesses occur with some uniform probability normalized to 

\begin{equation}
    \sum_{e}^{l-1}P(e|s)=1-p;\hspace{1em} s\neq e.
\end{equation}
    
We describe the agent's posterior of all bettors with the same form, with the "correct" bet binomial coefficient $x$.
Thus, calculating the growth rate becomes an expectation calculation over the set of received tips, summing over diagonal and off-diagonal components separately. 

The mutual information separates into a term of only $l=1/P(e)$, an on-diagonal, and off-diagonal term
\begin{equation}\label{appinf}
\begin{split}
    I(E;S)&=\sum_{e,s}^lP(e,s)\big[\log l+\log P(e|s)\big]\\
    &=\log l +p\log p +(1-p)\log \frac{1-p}{l-1}\\
    &=H(E)-H(E|S),
\end{split}
\end{equation}
with the entropy of the outcome given by $H(E)=\log l$ and the reduction in entropy by the signal given by  $H(E|S)=-p \log p -(1-p) \log \frac{1-p}{l-1}$. 
The information maximizes as $p\rightarrow 1$ and increases with $l$.
The information vanishes at $p\rightarrow 1/l$.
The divergence is 
\begin{equation}
\begin{split}
    \textrm E_s\big[D_{KL}(P||X)\big]&=\sum_{e,s}P(e,s)\log\frac{P(e|s)}{X(e|s)} \\
    &=p \log \frac{p}{x} +(1-p)\log \frac{1-p}{1-x},
    \end{split}
\end{equation}

\noindent which is always nonnegative and vanishes when $x\rightarrow p$. We can write the growth rate as the difference between these two terms
\begin{equation}\label{KellyG}
\gamma=\textrm{E}\big[\log lf(x,l)\big]=\log l + p\log x+(1-p)\log\frac{1-x}{l-1}.
\end{equation}

\subsection{Variance of growth model}\label{simplevariance}

The volatility can be calculated as the second moment of growth. Standard deviation is computed with the equation

\begin{equation}
    \sigma=\sqrt{E\big[\log \big(lf(x,l)\big)^2]-E\big[\log lf(x,l)\big]^2}.
\end{equation}

\noindent $\textrm E\big[\log lf(x,l)\big]$ is simply $\gamma$, and the second term is

\begin{equation}
\begin{split}
\textrm E\big[\log lf(x,l)\big]^2&= \bigg(\log l + p\log x+(1-p)\log\frac{1-x}{l-1}\bigg)^2 \\
&=\log^2l+p^2\log^2x+(1-p)^2\log^2\frac{1-x}{l-1} \\
    &+2p\log l\log x+2(1-p)\log l\log \frac{1-x}{l-1} \\
    &+2p(1-p)\log x\log\frac{1-x}{l-1}.
\end{split}
\end{equation}

\noindent The first term expands to

\begin{equation}
\begin{split}
    \textrm E\big[\log \big(lf(x,l)\big)^2\big]&=\textrm E_{e,s}\big[(\log P(e|s)+\log l\big)^2\big] \\
    &=\log^2l+p\log^2x+(1-p)\log^2\frac{1-x}{l-1} \\
    &+2p\log l\log x+2(1-p)\log l\log \frac{1-x}{l-1}.
\end{split}
\end{equation}

\ Combining these two quantities yields the volatility, where $(1-p)-(1-p)^2=p(1-p)$,

\begin{equation}
\begin{split}
\sigma_n&=\sqrt{p(1-p)\bigg[\log^2x+\log^2\frac{1-x}{l-1}-2\log x\log \frac{1-x}{l-1}\bigg]}\\
     &=\sqrt{p(1-p)}\log\frac{x(l-1)}{1-x}.
\end{split}
\end{equation}

\noindent The variance of investment clusters of size $1/\omega$ scales as 

\begin{equation}
    \sigma_t^2=\frac{1}{\gamma}\sigma_n^2,
\end{equation}

\noindent where the subscript $t$ denotes the temporal variance.

\section{Latent Dirichlet Allocation}

\subsection{Defining the model}\label{LDA}

In this section, we derive the Latent Dirichlet Allocation (LDA) mode for the degenerate multinomial environment. The Bayesian update equation is given by 

\begin{equation}\label{LDAupdate}
\begin{split}
 X(e|s)
 &\propto \frac{\big(m_{(-s)}^{(-e)}+\tilde\beta_s^e\big)}{(M^{(-s)}+\tilde B^e)} (n_{(-e)}+\tilde\alpha_e), 
\end{split}
\end{equation}

\noindent for $m^{(-e)}_{(-s)}$ occurrences of $s$ conditional on $e$ excluding the current, $n_{(-e)}$ occurrences of $e$ excluding the current in a batch of $n$ trials.
We set  $\alpha_e=1$, as every event is equally likely. 
For $s=e$, $\beta_{es}=x$, and for $s\neq e$, $\beta_{se}=\frac{(l-1)}{1-x}$ to impose degenerate off-diagonal conditions on $s|e$.
We introduce $B_s=\sum_e\beta^e_s$, whereby symmetry, $B_s\equiv B=1$, and we count over the diagonals, $n_{e=s}$, and off diagonals, $n_{e\neq s}$. 
Therefore the diagonal environmental posterior is

\begin{equation}
 P(e|s) \propto\frac{\big(m_{(-s=e)}^{(-e)}+x_e\big)}{\big(M^{(-s)}+1\big)}(n_{(-e)}+1),
\end{equation}

\noindent and the off-diagonal is
\begin{equation}
 P(e|s)\propto\frac{\big(m_{(-s\neq e)}^{(-e)}+\frac{1-x_e}{l-1}\big)}{\big(M^{(-s)}+1\big)} (n_{(-e)}+1).
\end{equation}

\subsection{Asymptotic, temporal behavior}\label{asympt}

We introduce the temporal behavior, with two constants. 
We multiply the number of observations by the observation rate $\omega$, with units $samples/time$ and the inference rate $k$, with unit $time/update$.
The inference rate counts the number of samples per Bayesian update, and the observation rate counts the updates per unit time. 
We multiply through by $k$ so that it becomes a magnitude on the hyperprior, leaving 

\begin{equation}\label{app:LDAupdate}
\begin{split}
 P(e|s)
 &\propto \frac{(m_{(-s)}^{(-e)}/\omega+\tilde \beta_s^ek)}{(M^{(-s)}/\omega+1k)} (n_{(-e)}/\omega+\tilde 1/k).
\end{split}
\end{equation}

Over many observations, the law of large numbers argues that each outcome count converges to the environmental posterior with some noise, $\xi_i$ as

\begin{equation}
\begin{split}
M^{(-s)}/\omega&\rightarrow P(s)Nt+\xi_s \\
n_{(-e)}/\omega&\rightarrow P(e)Nt+\xi_e \\
m_{(-s)}^{(-e)}/\omega&\rightarrow P(s|e)Nt+\xi_{s|e},
\end{split}
\end{equation}
where the $\xi'$s are fluctuation terms representing deviations from the mean. 
Over many i.i.d observations of events, $\xi\rightarrow 0$. 
The marginal terms converge to uniform over all states and become unity, and the agent posterior converges to the dynamical distribution (APP 5)

\begin{equation}\label{simplelearn}
    X(e,\lambda|s)= \frac{P(s|e) \lambda+X(s|e)}{1+ \lambda },
\end{equation}

\noindent where we have converted to the time domain $t=N/\omega$, and substituted the unitless inference sample size $ \lambda=t/kl$.
Over long times, the distribution converges to the environmental posterior by

\begin{equation}
\begin{split}
X(e,\lambda|s)&\propto\frac{P(s|e) \lambda+X(s|e)}{P(s) \lambda+1}\big(P(e) \lambda+\alpha_e\big)\\ \\
&\rightarrow \bigg(P(s|e)+\frac{X(s|e,0)}{ \lambda}\bigg)\frac{P(e)}{P(s)}=P(e|s),
\end{split}
\end{equation}

\noindent yielding power law time-averaged behavior. 
At early times, as $t\rightarrow 0$ the posterior is proportional to the agent's initial agent posterior, $X(E|S)$, and converges to $P(E|S)$ as $kl\ll t\rightarrow\infty$.
If agents are initialized with the same diagonal posterior value such that $X(s|e)=X(e^\prime,s^\prime)$ for all $e=e,s^\prime=s^\prime$, we can assume that the diagonals of an agent uniformly converge to $p$ in time such that $X(s,t|e)\equiv x(t)$ for all $s=e$,  

\subsection{Growth rate variance}\label{GRV}

The mean growth rate is computed, where for brevity, the expected divergence for agent $i$ with signals $s_i\in S_i$ is given as $\mathrm{E}_{s_i}\big(D_{KL}\big[P(E|s_i)||X(E|s_i\big]\big)\equiv D_i$, and the mutual information between individual signals and the environment, $I(E;S_i)\equiv I_i$

\begin{equation}
    \begin{split}
    \langle\gamma\rangle&=\frac{1}N\sum_iI(E|S_i)-\textrm{E}_{s_i}\big[D_i\big]\\
        &=\big\langle I\big\rangle-\langle D\rangle,
    \end{split}
\end{equation}

\noindent where angle brackets denote population arithmetic means. The variance in growth rates is calculated

\begin{equation}
    \begin{split}
        \textrm{Var}_N[\gamma_i]&=\big\langle(\gamma-\langle\gamma\rangle)^2\big\rangle, \\
        &=\big\langle\gamma^2+\langle\gamma\rangle^2-
        2\gamma\langle\gamma\rangle\big\rangle\\
        &=\langle I^2\rangle-\langle I\rangle^2+\langle D^2\rangle-\langle D\rangle^2 \\
        &-2\big(\langle ID \rangle-\langle I\rangle\langle D \rangle\big)\\
        &=\textrm{Var}_N[I_i]+\textrm{Var}_N[D_i]-2\textrm{Covar}_N\big[I_iD_i\big].
    \end{split}
\end{equation}

\noindent When all agents are exposed to the same environment, the first and third terms vanish, leaving

\begin{equation}
        \textrm{Var}_N[\gamma_i]=\textrm{Var}_N\bigg[\textrm{E}_{s_i}\bigg(D_{KL}\big[P(S|s_i)||X(E|s_i)\big]\bigg)\bigg].
\end{equation}

\subsection{Binomial parameter variance }

The binomial variance can be computed exactly as

\begin{equation}
\begin{split}
      \textrm{Var}_N\big[x_i(\lambda)\big]&=\frac{1}{N }\sum_i^N \bigg[\frac{p \lambda+x_i}{1+ \lambda}\bigg]^2-\bigg[\frac{p \lambda+\langle x\rangle}{1+ \lambda}\bigg]^2 \\
      &=\frac{1}{N }  \sum_i^N \bigg[2\frac{x_ip \lambda+x_i^2}{(1+ \lambda)^2}-2\frac{\langle x\rangle p \lambda-\langle x\rangle^2}{(1+ \lambda)^2}\bigg] \\
      &= \frac{\langle x^2\rangle-\langle x\rangle^2}{ (1+ \lambda)^2}=\frac{\sigma_x^2}{(1+ \lambda)^2}.
\end{split}
\end{equation}

\subsection{Multinomial growth rate variance }
\label{growthVar} 

The variance of a function, $\gamma(x)$, of a random variable, $x$, is given generally by the Taylor expansion of that function \cite{5790}.
It is written as

\begin{equation}
\begin{split}
    \textrm{Var}_N\big(\gamma[x_i(\lambda)])&=\gamma^\prime\big[\big\langle x(\lambda)\big\rangle\big]\textrm{Var}_N\big[x_i(\lambda)\big]\\
    &-\frac{\gamma^{\prime\prime}\big[\big\langle x(\lambda)\big\rangle\big]^2}{4}\textrm{Var}_N^2\big[x_i(\lambda)\big]+\bar T^3,
\end{split}
\end{equation}

\noindent where primes denote differentiation with respect to $x$, and $\bar T^3$ are higher order terms that are only relevant at small times. 
The first and second-order derivatives of $\gamma$ are given by 

\begin{equation}
\begin{split}
\gamma^\prime(x)&=\frac{p}{x}-\frac{1-p}{1-x} \\
\gamma^{\prime\prime}(x)&=-\bigg[\frac{p}{x^2}+\frac{1-p}{(1-x)^2}\bigg],
\end{split}
\end{equation}

\noindent and the variance term is given by

\begin{equation}
\textrm{Var}_N\big[x_i(\lambda)\big]=\frac{\sigma_x^2}{(1+\lambda)^2}.
\end{equation}

\noindent The growth rate variance after small times is given by

\begin{equation}\label{gvar}
\begin{split}
    \textrm{Var}_N\big (\gamma[x_i(\lambda)]\big)&=\bigg[\frac{p}{\bar x}-\frac{1-p}{1-\bar x}\bigg]\frac{\sigma_x^2}{(1+\lambda)^2}\\
    &+\bigg[\frac{p}{\bar x^2}+\frac{1-p}{(1-\bar x)^2}\bigg]\bigg[\frac{\sigma_x^2}{(1+\lambda)^2}\bigg]^2,
\end{split}
\end{equation}

\noindent were for brevity, $\bar x\equiv\big\langle x(\lambda)\big\rangle$. 

\end{document}